\documentclass[useAMS,usenatbib]{mn2e}

\usepackage{graphicx}
\usepackage{color}

\title[Type Ic supernovae from very massive stars]{Type Ic core-collapse supernova explosions evolved from very massive stars}
\author[T. Yoshida, S. Okita, and H. Umeda]
{Takashi Yoshida$^{1}$\thanks{E-mail: yoshida@yukawa.kyoto-u.ac.jp}, 
Shinpei Okita$^{2}$ and Hideyuki Umeda$^{2}$\\
$^{1}$Yukawa Institute for Theoretical Physics, Kyoto University, Kyoto 606-8502, Japan\\
$^{2}$Department of Astronomy, Graduate School of Science, University of Tokyo, 
Tokyo 113-0033, Japan
}
\begin{document}

\date{Accepted 2013 December 12. Received 2013 December 9; in original form 2012 December 25}

\pagerange{\pageref{firstpage}--\pageref{lastpage}} \pubyear{2013}

\maketitle

\label{firstpage}

\begin{abstract}
We investigate the possibility of a super-luminous Type Ic 
core-collapse supernovae producing a large amount of $^{56}$Ni.
Very massive stars with a main-sequence mass larger than 100 M$_\odot$ and 
a metallicity $0.001 < Z \la 0.004$ are expected to explode as super-luminous Type Ic supernovae.
Stars with $\sim 110 - 150$ M$_\odot$ and $Z \la 0.001$ would explode as Type Ic
pulsational pair-instability supernovae if the whole H and He layers has been lost by the mass loss
during pulsational pair-instability.
We evaluate the total ejecta mass and the yields of $^{56}$Ni, 
O, and Si in core-collapse supernovae evolved from very massive stars.
We adopt 43.1 and 61.1 M$_\odot$ WO stars with $Z=0.004$ as supernova progenitors
expected to explode as Type Ic core-collapse supernovae.
These progenitors have masses of 110 and 250 M$_\odot$ at the zero-age main sequence.
Spherical explosions with an explosion energy larger than 
$2 \times 10^{52}$ erg produce more than 3.5 M$_\odot$ 
$^{56}$Ni, enough to reproduce the light curve of SN 2007bi.
Asphericity of the explosion affects the total ejecta mass as well as the yields of $^{56}$Ni,
O, and Si.
Aspherical explosions of the 110 and 250 M$_\odot$ 
models reproduce the $^{56}$Ni yield of SN 2007bi.
These explosions will also show large velocity dispersion.
An aspherical core-collapse supernova evolved from a very massive star is a possibility
of the explosion of SN 2007bi.
\end{abstract}


\section{Introduction}

Recent supernova (SN) surveys found a variety of super-luminous SNe (SLSNe) 
in metal-poor galaxies \citep{Quimby11,Gal-Yam12}.
SLSNe indicate peak absolute magnitude less than $\sim -21$ mag and
have a diversity of the rise and decline time in their light curves.
SN 2007bi is a Type Ic SLSN \citep{Gal-Yam09}.
Its light curve is well fitted  to the radioactive decay of $^{56}$Co, so that this 
explosion is considered to be powered by large amount of $^{56}$Ni produced during 
the explosion of a very massive star \citep[SLSN-R in][]{Gal-Yam12}.
Since observational analyses indicated that the $^{56}$Ni yield is $3.5 - 7.4$ M$_\odot$, 
possibilities of pair-instability (PI) SN and core-collapse (CC) SN were proposed as 
the explosion mechanism \citep{Gal-Yam09,Moriya10}.
The range of the main-sequence (MS) mass appropriate for SN 2007bi was evaluated as 
$525 - 575$ M$_\odot$ for PI SN and $110 - 270$ M$_\odot$ for CC SN explosions, 
\citep[hereafter abbreviated by YU11]{Yoshida11}, in $Z = 0.004$ stars.

Asphericity of energetic CC explosion has been discussed as a comparison with 
the abundance patterns of extremely metal-poor (EMP) stars \citep[e.g.][]{Umeda02,Umeda05}.
Aspherical hypernovae well reproduced the abundance pattern of EMP stars \citep{Tominaga09},
so that hypernova explosion is expected to be aspherical.
On the other hand, some energetic Type Ic SNe (SNe Ic) showed evidence for asphericity in late time spectra \citep{Maeda08}.
The observational or indirect evidence of aspherical energetic CC explosions also suggests a possibility of aspherical CC explosion for SN 2007bi.

%
\begin{figure*}
\includegraphics[width=6cm,angle=270]{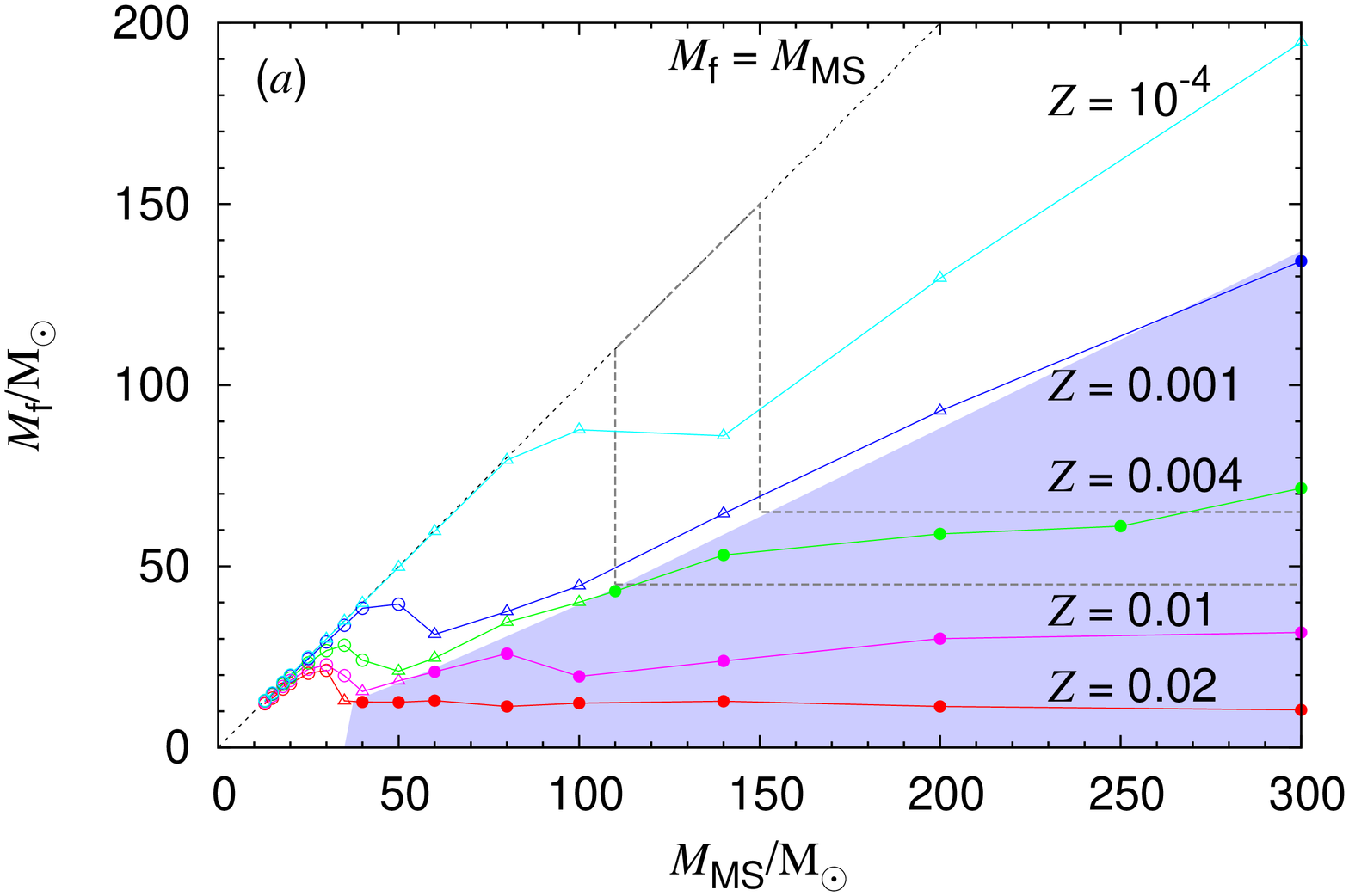}
\includegraphics[width=6cm,angle=270]{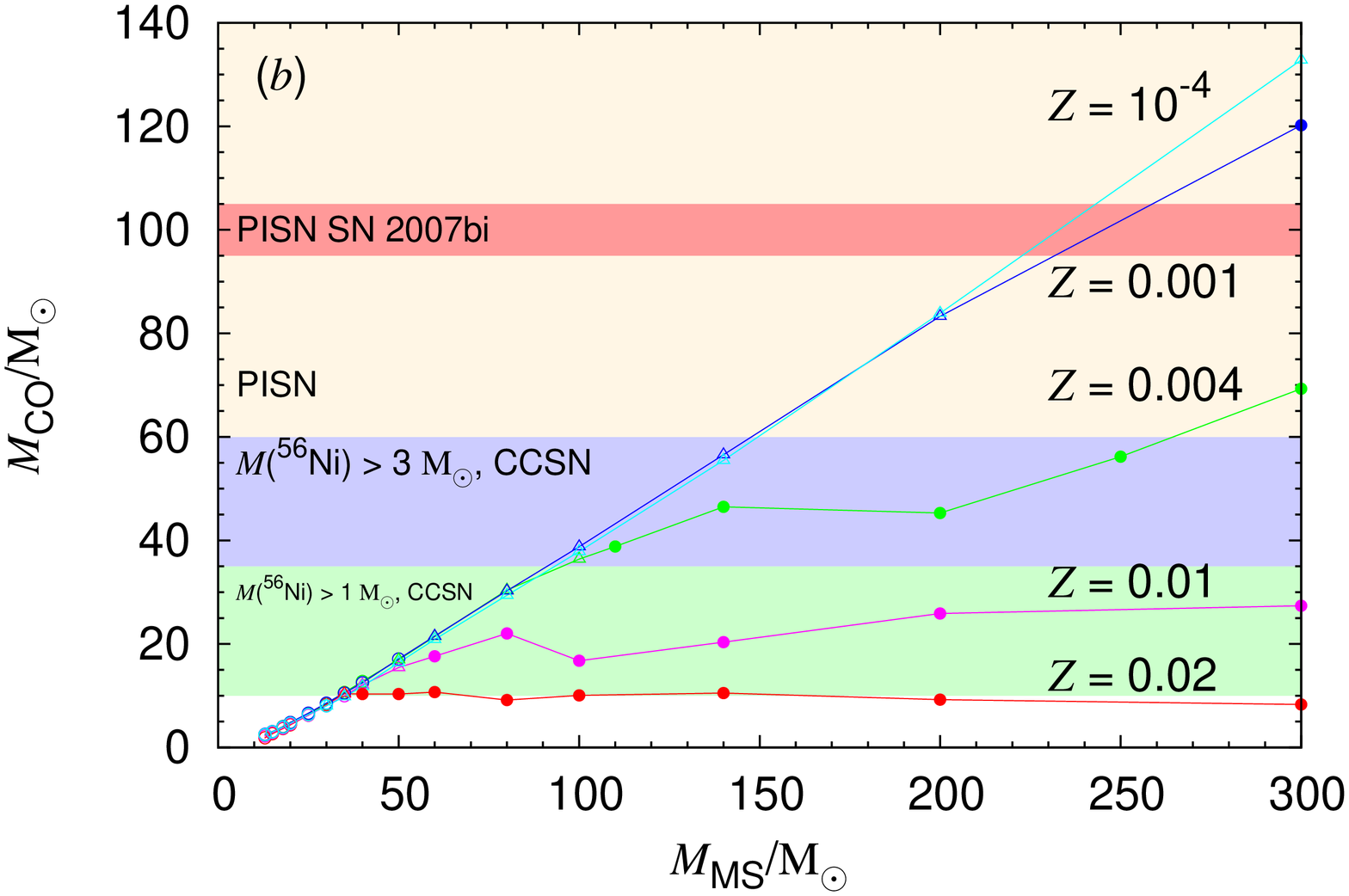}
\caption{The relation of the final mass $M_{{\rm f}}$ (panel ($a$)) and 
the CO core mass $M_{{\rm CO}}$ (panel ($b$)) to the MS mass $M_{{\rm MS}}$ 
of stars with metallicities
$Z$ = 0.02 (red line), 0.01 (purple line), 0.004 (green line), 0.001 (blue line), 
and $10^{-4}$ (cyan line).
Open circles, triangles, and closed circles indicate red supergiants,
yellow supergiants or WN stars, and WO stars, respectively.
In panel ($a$), progenitors in the blue shaded region are WO stars and are expected to 
explode as SNe Ic.
The stars in the region enclosed by the dashed line will experience PPI (see text in details).
The dotted line shows $M_{{\rm f}} = M_{{\rm MS}}$.
In panel ($b$), green and blue regions indicate the CC SN explosions ejecting
$^{56}$Ni more than 1 M$_\odot$ and 3 M$_\odot$, respectively.
Orange region indicates PISN.
Red region is the CO core mass range expected to reproduce the $^{56}$Ni yield in SN 2007bi.
}
\end{figure*}

Recent observations revealed the existence of very massive stars.
WN stars of $92 - 265$ M$_\odot$ were found in the young
star clusters NGC 3603 and R136 \citep{Crowther10}.
Their zero-age MS masses were evaluated up to 320 M$_\odot$.
The evolution of very massive stars has been investigated and the final
fates such as PI SNe and pulsational pair-instability (PPI) SNe were discussed 
\citep[e.g.,][]{Langer07,Waldman08,Yungelson08,Chatzopoulos12a,Yusof13}.
Recently, effects of eruptive mass loss of very massive stars by PPI
were also discussed \citep{Woosley07,Chatzopoulos12}.
Although mass loss rate of very massive stars is still uncertain, fates of 
very massive stars with current mass loss rate should be investigated.

In this paper, we investigate the dependence of the final mass, the CO core mass, and
the stellar types on the MS mass with the metallicity range
$Z=10^{-4} - 0.02$.
We discuss progenitors appropriate for CC SLSNe Ic.
We also investigate the explosive nucleosynthesis of CC SNe 
evolved from very massive stars.
We consider two different features of explosions: spherically
symmetrical explosions with different explosion energies and 2D
aspherical explosions with different opening angles of mass ejected
region.
We discuss the dependence of the total ejecta mass and the
yields of $^{56}$Ni, O, and Si on explosion features.

In Section 2, we show the final mass and the CO core mass among stars with
an MS mass $M_{{\rm MS}} = 13-300$ M$_\odot$ and a metallicity $Z=10^{-4}-0.02$.
We also show the progenitor models evolved from 110 M$_\odot$ 
and 250 M$_\odot$ stars with $Z = 0.004$ as progenitors of SLSNe Ic.
In Section 3, we show the dependence of the yields of $^{56}$Ni, O,
and Si on the explosion energy of spherically symmetrical CC SN explosions.
In section 4, we present the dependence on asphericity of 
aspherical CC SN explosions.
Progenitors of SLSNe Ic, especially SN 2007bi, and aspherical CC SNe Ic evolved from 
very massive stars are discussed in Section 5.
Conclusions are presented in Section 6.

\section[]{Very Massive Progenitors}

We calculate the evolution of massive stars until the central
C-depletion with the ranges of an MS mass $M_{{\rm MS}} = 13 - 300$ M$_\odot$ and 
a metallicity $Z = 10^{-4} - 0.02$ using the stellar
evolution code in YU11 and \citet{Umeda12}.
The species of nuclei adopted in the nuclear reaction network is listed in Table 1.
The mass loss recipe taken in this study is the same as in Case (A) in YU11.

\begin{table}
\caption{
Nuclear reaction network used for stellar evolution and explosive nucleosynthesis during SN explosion. Isomer of $^{26}$Al is taken into account.}
\begin{center}
\begin{tabular}{lc|lc}
\hline
Element & $A$ & Element & $A$ \\
\hline
n & 1 & Ar & $34-43$ \\
H & $1-3$ & K & $36-45$ \\
He & $3-4$ & Ca & $38-48$ \\
Li & $6-7$ & Sc & $40-49$ \\
Be & $7,9$ & Ti & $42-51$ \\
B & $8,10,11$ & V & $44-53$ \\
C & $11-13$ & Cr & $46-55$ \\
N & $13-15$ & Mn & $48-57$ \\
O & $14-18$ & Fe & $50-61$ \\
F  & $17-19$ & Co & $51-62$ \\
Ne & $18-22$ & Ni & $54-66$ \\
Na & $21-23$ & Cu & $56-68$ \\
Mg & $22-27$ & Zn & $59-71$ \\
Al & $25-29$ & Ga & $61-73$ \\
Si & $26-32$ & Ge & $63-75$ \\
P & $27-34$ & As & $65-76$ \\
S & $30-37$ & Se & $67-78$ \\
Cl & $32-38$ & Br & $69-79$ \\
\hline
\end{tabular}
\end{center}
\end{table}

Figure 1($a$) shows the relation between the MS mass and the final mass of
massive stars.
In stars with $M_{{\rm MS}} \ga 40$ M$_\odot$ and $Z \ga 0.01$, strong
mass loss strips the whole H and/or He layers. These stars 
become Wolf-Rayet (WR) stars 
and their final masses become $\sim 10 - 30$ M$_\odot$.
Very massive stars with $M_{{\rm MS}} \ga 100$ M$_\odot$ and $Z=0.004$ become WO stars.
Stars with $Z \la 0.001$ possess the H and/or He layers until the C-burning.
If the whole H and He layers is not lost by eruptive mass loss induced by luminous blue
variable-like events or PPI, these stars would explode as SNe II or Ib.
We will discuss the effect by eruptive mass loss during PPI later.

\begin{table}
\caption{
Progenitors of SNe Ic.
}
\begin{center}
\begin{tabular}{lcc}
\hline
$M_{{\rm MS}}$/M$_\odot$ & 110 & 250 \\
\hline
$M_{{\rm f}}$/M$_\odot$         & 43.1 & 61.1  \\
$M_{{\rm CO}}$/M$_\odot$ & 38.2  & 56.2  \\
$M_{{\rm Fe}}$/M$_\odot$ & 3.03  & 3.21  \\
$M({\rm He})$/M$_\odot$ & 0.24 & 0.51  \\
$Y_S$      & 0.191  & 0.193 \\
\hline
\end{tabular}
\end{center}
\end{table}

Figure 1($b$) shows the relation between the CO core mass and the MS mass.
The maximum mass of the CO core increases with decreasing the metallicity.
Stars with $Z=0.01$ and $Z=0.004$ have larger CO core than stars
with the same MS mass and $Z=0.02$ owing to smaller mass loss rate.
The mass range of CO core among WR stars depends on metallicity.
The ranges are $8-11$  M$_\odot$, $17-27$ M$_\odot$, and $39-69$ M$_\odot$
for WO stars with $Z=0.02, 0.01$, and 0.004, respectively.
Among stars with $Z=0.001$ and $10^{-4}$, metallicity dependence of the CO
core mass is quite small.

We note that stars having a CO core of $\sim 40 - 60$ M$_\odot$ experience PPI \citep{Heger02} 
and outer layer is lost \citep{Woosley07}.
We expect that the stars in the region enclosed by the dashed line in Fig. 1($a$) will become PPI.
The ejected amount of the outer layer will strongly depend on pulsations of PPI and the density
structure.
If the whole H and He layers of stars with $M_{{\rm MS}} \sim 110 - 150$ M$_\odot$ and 
$Z \la 0.001$ is lost by eruptive mass loss during PPI, these stars will explode as SNe Ic.

In the following sections, we use progenitors to study SLSNe Ic 
evolved from very massive stars with MS masses 
$M_{{\rm MS}} = 110$ M$_\odot$ and 250 M$_\odot$ and a metallicity $Z = 0.004$.
Since the metallicity of the host galaxy of SN 2007bi is $Z= 0.2-0.4 Z_\odot$
\citep{Young10}, the metallicity of our progenitor models is within the
range of the host galaxy of SN 2007bi.
We calculated the evolution of these two models up to the onset of the core collapse, i.e., 
the central temperature of $\log T_{\rm C} \sim 9.8$.
We added an acceleration term to the hydrostatic equation after the C-burning
\citep[e.g.,][]{Sugimoto81}.
These models lose all H-rich envelope during He burning and almost all He layer after the central He burning and evolve to WO stars.
The final mass $M_{\rm f}$, the CO core mass $M_{\rm CO}$, the Fe core mass $M_{\rm Fe}$,
the surface He amount $M({\rm He})$, and the He mass fraction at the surface $Y_S$ of
these progenitors are listed in Table 2.
The CO core is defined as the region where He mass fraction is 
smaller than $10^{-3}$ \citep{Umeda08}.
The Fe core is defined as the region where the total mass fraction
of the elements heavier than Sc is larger than 0.5 \citep{Hirschi04}.
These stars are expected to explode as a SN Ic because of small He mass fraction at surface
\citep[e.g.,][]{Yoon10}.
The $M_{{\rm MS}} = 250$ M$_\odot$ star experienced PPI during the
Si burning \citep[e.g.,][]{Umeda08}.
We do not consider eruptive mass loss during PPI.
The effect of the eruptive mass loss will be discussed in \S 5.

\begin{figure}
\includegraphics[width=6cm,angle=270]{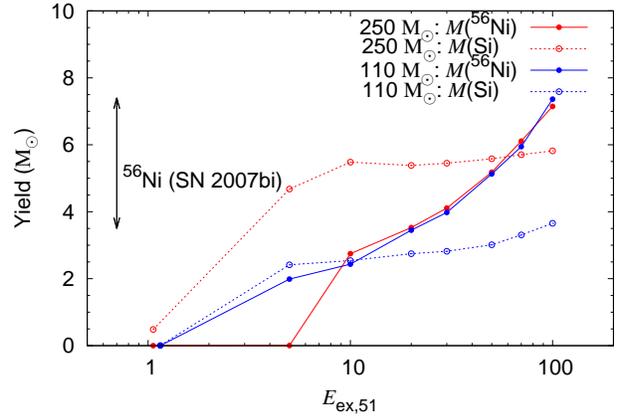}
\caption{The yields of $^{56}$Ni (solid line with closed circles) and Si (dotted line with 
open circles) ejected by spherical explosions of the 110 and 250   M$_\odot$ models 
as a function of the explosion energy $E_{{\rm ex},51}$.
Red and blue lines indicate the yields of the 110 and 250 M$_\odot$ models, respectively.
The observed range of the $^{56}$Ni yield in SN 2007bi is indicated
by the vertical arrow.
}
\end{figure}

\section{Spherical Supernova Explosion}

First, we investigate the explosive nucleosynthesis of spherical SN explosions.
We calculate time evolution of 1D spherical SN explosion using a PPM code
\citep{Colella84} as in \citet{Umeda05}.
Thermal energy is injected inside the mass cut to explode the star.
We set the location of the mass cut to be 2.0 M$_\odot$.
We consider eight explosion models of which explosion energies 
above the mass cut are
$E_{{\rm ex},51} = 1, 5, 10, 20, 30, 50, 70$, and 100 where 
$E_{{\rm ex},51}$ is the explosion energy in units of $10^{51}$ erg.
Then, we calculate the nucleosynthesis during the SN explosions
by post-processing.
The nuclear reaction network is the same as in the stellar evolution code.
When the temperature is higher than $9 \times 10^9$ K, chemical
composition is solved assuming nuclear statistical equilibrium.
The $\nu$-process is included with a  parameter set independent
of the explosion model as in \citet{Yoshida08}.
The total neutrino energy is assumed to be $3 \times 10^{53}$ erg.
The neutrino luminosity decreases exponentially with time with a time scale of 3 s.
The neutrino spectra obey Fermi-Dirac distributions with zero chemical potentials.
The temperatures are set to be 4.0 MeV for $\nu_e$ and $\bar{\nu}_e$ and
6.0 MeV for $\nu_{\mu,\tau}$ and $\bar{\nu}_{\mu,\tau}$.
The rates of the $\nu$-process reactions are adopted from
Hoffman \& Woosley (1992; unpublished)\footnote{http://ie.lbl.gov/astro/hw92\_1.html},
\citet{Horowitz02}, \citet{Yoshida08b}, and \citet{Suzuki09}.
We note that although the $\nu$-process has large uncertainties in
SN explosions of very massive stars, it is not important for the
production of $^{56}$Ni, O, and Si.

Figure 2 shows $^{56}$Ni and Si yields in the SN ejecta as a function of the explosion 
energy $E_{{\rm ex},51}$.
The obtained $^{56}$Ni yield will correspond to the maximum yield produced through 
aspherical explosion with the same explosion energy.
The $^{56}$Ni yield increases with the explosion energy and scarcely depends on the
mass of the progenitor in the case of $E_{{\rm ex,51}} \ga 10$.
The maximum $^{56}$Ni yield is 6.8 M$_\odot$ and 6.4 M$_\odot$ in 
$E_{{\rm ex},51} = 100$ for the 110 and 250 M$_\odot$ models.
The $^{56}$Ni yield of more than 3.5 M$_\odot$, which reproduces the amount observed in 
SN 2007bi, is obtained from the explosions with $E_{{\rm ex,51}} \ga 20$.
The Si yield of the 110 and 250 M$_\odot$ SN models are 
$M({\rm Si}) \sim 2.45 - 3.65$ M$_\odot$ and $5.38 - 5.82$ M$_\odot$ 
in $E_{{\rm ex,51}} \ga 10$.
The explosion-energy dependence is smaller than that of $^{56}$Ni.
As the explosion energy increases, the Si abundant region shifts outward
but the width of the Si abundant region in the mass coordinate scarcely 
changes.
The Si yield of the 250 M$_\odot$ model is larger than that in the 110 M$_\odot$ model.
More Si is produced through the O shell burning during and after PPI in the 250 M$_\odot$ model.

We note that some materials fell-back to the central remnant in the explosions with
$E_{{\rm ex,51}} = 1$ for the 110 M$_\odot$ model and with 
$E_{{\rm ex,51}} = 1$ and 5 for the 250 M$_\odot$ model.
In these three models, the baryon masses of the final remnants are 28.17 M$_\odot$,
46.56 M$_\odot$, and 7.32 M$_\odot$.
Although about 2 M$_\odot$ of $^{56}$Ni was synthesized during the explosions, almost
all $^{56}$Ni fell-back.
Therefore, these explosions are not classified into SLSNe.

\section{Aspherical Supernova Explosion}

We calculate aspherical SN explosions with different opening angles 
using 2D Eulerian aspherical hydrodynamic code 
\citep[see Okita \& Umeda 2013, in preparation, for details; see also][]{Okita12}.
We initially inject kinetic energy for $10^{-3}$ s in the innermost region of the ejecta
to explode with a total explosion energy $E_{{\rm ex,51}} = 50$ and 70 in the cases
of the 110 M$_\odot$ model and the 250 M$_\odot$ model, respectively.
These explosion energies in spherical explosions roughly reproduce the photometric velocity of 
SN 2007bi $\sim 12,000$ km s$^{-1}$ identified by \citet{Gal-Yam09}.
We discuss ejecta velocities in aspherical explosions later.
The injected kinetic energy is immediately converted into local thermal energy.
We locate a central remnant of 2.0 M$_\odot$ corresponding to
the mass cut for spherical explosion.
An approximated analytical equation of state including ultra-relativistic electrons and positrons
 at high temperature  \citep[e.g.,][]{Tominaga09} is adopted to the calculations;
\begin{equation}
P = \frac{kT}{\mu m_u} + \frac{a T^4}{3} \left( 1+\frac{7}{4} \frac{T_9^2}{T_9^2+5.3} \right),
\end{equation}
where $P$ is pressure, $k$ is Boltzmann constant, $T$ is temperature, 
$\mu$ is the mean molecular weight, which is assumed to be 2 here, 
$m_u$ is the atomic mass unit, 
$a$ is a radiation density constant, and $T_9$ is temperature 
in units of $10^9$ K.
We choose several different opening angles $\theta_{op}$ of the polar ejected region.
Then, we calculate the nucleosynthesis in the SN ejecta post-processingly
using particle trace method.
The orbits of 5,200 Lagrangian particles are taken into account.

\begin{figure}
\includegraphics[width=6cm,angle=270]{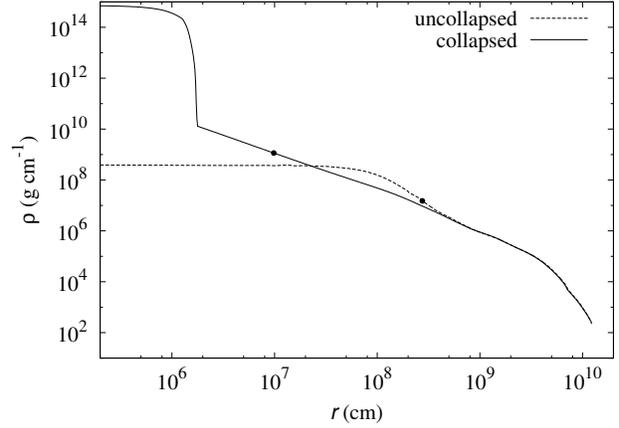}
\caption{
Density distribution as a function of the radius for the uncollapsed (dashed line)
and collapsed (solid line) progenitors of the 110 M$_\odot$ model.
The points indicate the location of the mass cut at 2.0 M$_\odot$ in the mass coordinate.
}
\end{figure}

\begin{figure*}
\includegraphics[width=6cm,angle=270]{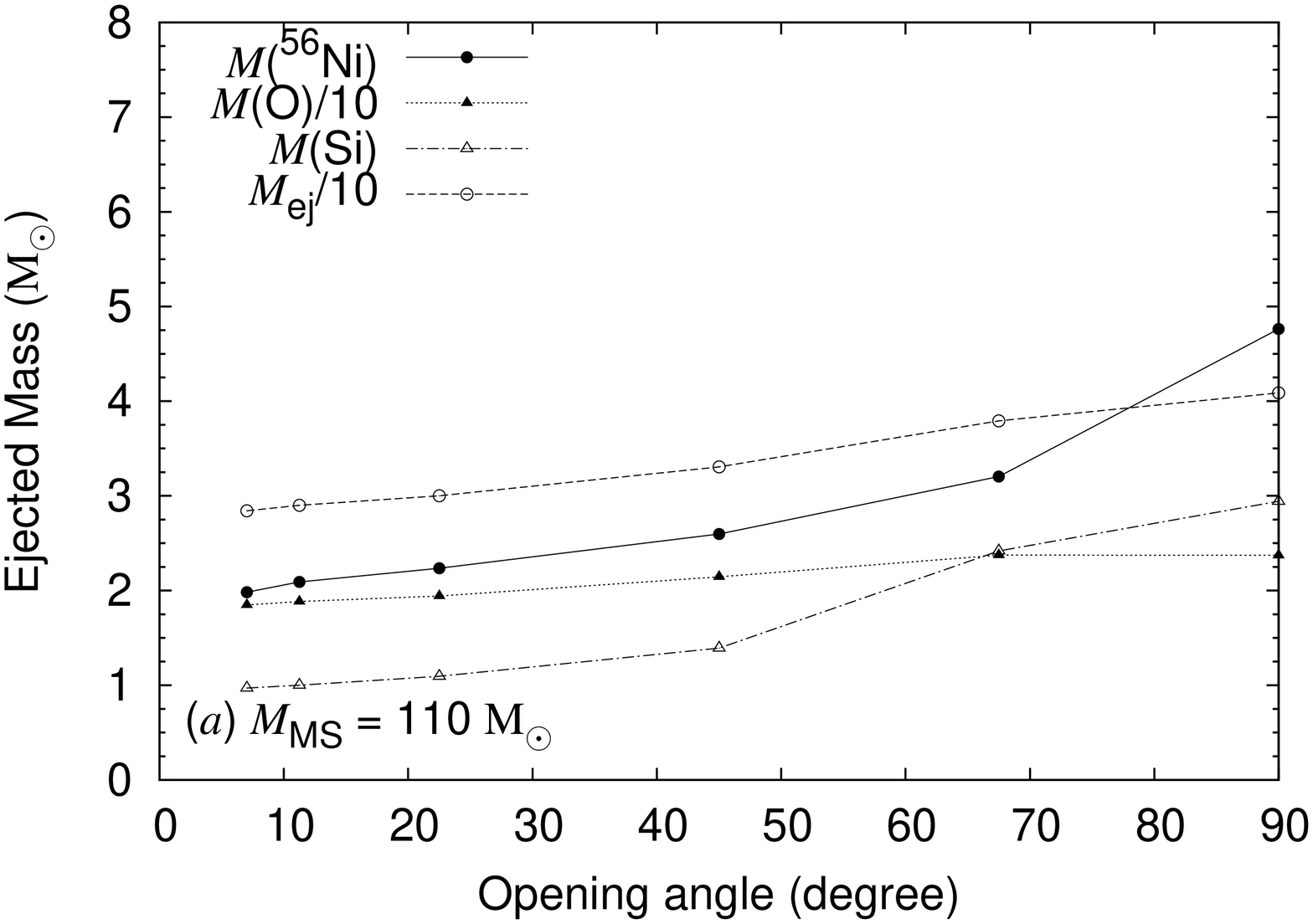}
\includegraphics[width=6cm,angle=270]{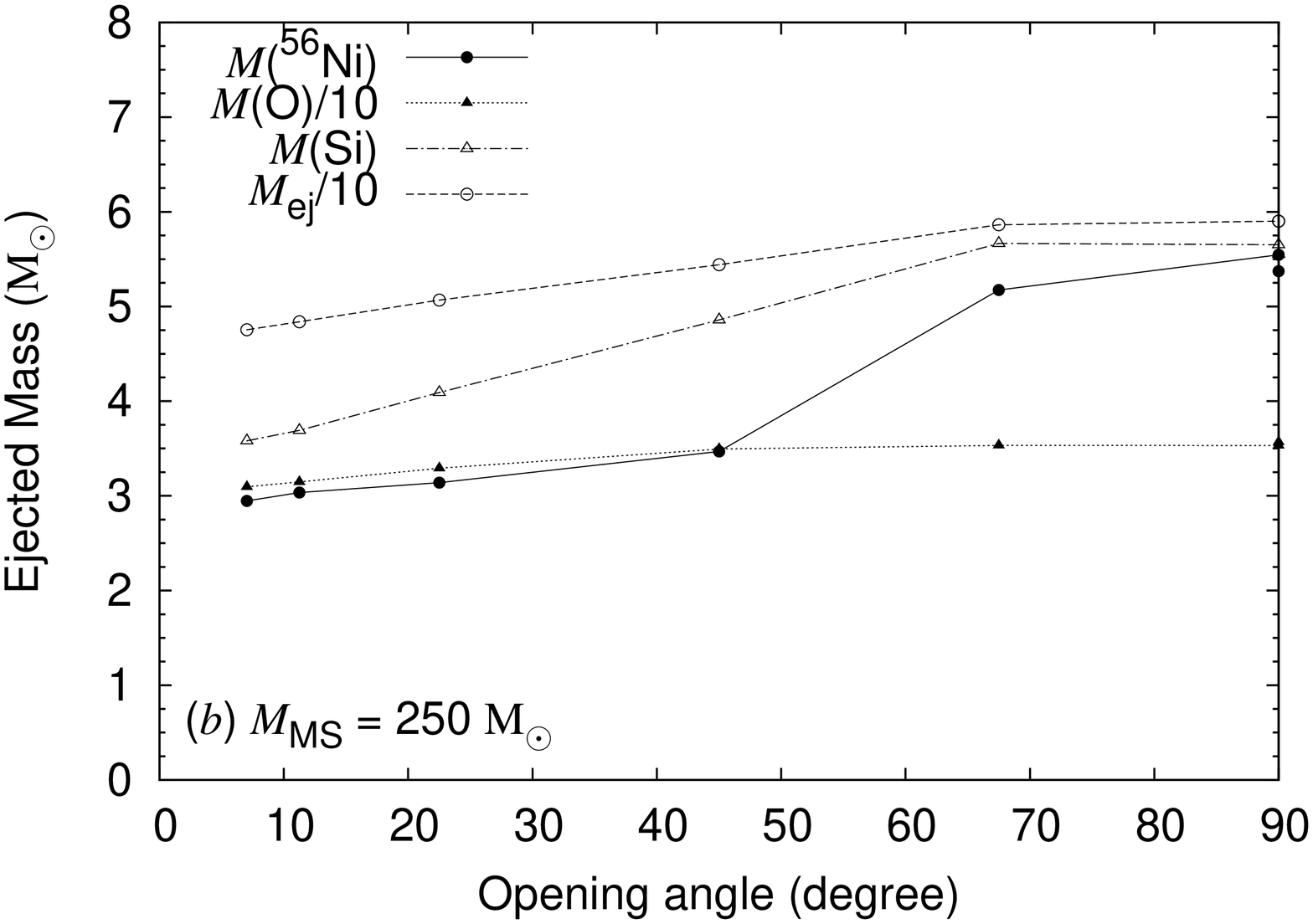}
\caption{The yields of $^{56}$Ni (solid line), O (dotted line),
and Si (dash-dotted line), and the total ejecta 
amount $M_{\rm ejecta}$ (dashed line) in the aspherical explosion models of
the uncollapsed progenitors.
The MS mass $M_{{\rm MS}}$ and the explosion energy $E_{{\rm ex,51}}$ are 
set to be ($a$) 110 M$_\odot$ and 50 and ($b$) 250 M$_\odot$ and 70.
}
\end{figure*}

In this study, we adopt two density profiles for each progenitor as the initial conditions
of the hydrodynamical calculations.
The first one is the final density profile of the stellar evolution calculation.
The other is the density profile obtained by the core-collapse calculation.
We pursued the core-collapse until the mass of the collapsed matter corresponding to
a proto-neutron star or a black-hole becomes about 2 M$_\odot$.
During the core-collapse, the radius at the mass coordinate of 2.0 M$_\odot$ changed from
$2.9 \times 10^8$ cm to $9.8 \times 10^6$ cm for the 110 M$_\odot$ model and from
$1.5 \times 10^8$ cm to $1.0 \times 10^7$ cm for the 250 M$_\odot$ model.
We call the former progenitor ^^ ^^ uncollapsed progenitor model" and the latter progenitor
^^ ^^ collapsed progenitor model".
Figure 3 shows the density distribution of the uncollapsed and collapsed
progenitors of the 110 M$_\odot$ model.
Since the central region of the progenitor should contract before aspherical explosion, 
we consider that collapsed progenitor model shows more realistic feature.

\subsection{Aspherical explosions from uncollapsed progenitors}

We first evaluate the yields of the aspherical SN explosions from the uncollapsed 110 and
250 M$_\odot$ models.
Figure 4 shows the yields of $^{56}$Ni, O, and Si and the total ejecta mass as a function of
the opening angle.
The $^{56}$Ni yield ranges in $2.0 - 4.8$ M$_\odot$ for
the 110 M$_\odot$ model and 
$2.9 - 5.5$ M$_\odot$ for the 250 M$_\odot$ model.
The yield in small opening angle with 
$\theta_{op} \la 45^\circ$ for 110 M$_\odot$
model is about $40 \%$ of the yield for the spherical explosion.
The aspherical explosion of the 250 M$_\odot$ model with $\theta_{op} \la 45^\circ$ produces 
$^{56}$Ni of about a half amount compared with produced in the spherical explosion.
On the other hand, large opening angle case with $\theta_{op} \ga 68^\circ$ 
in the 250 M$_\odot$ model indicates the $^{56}$Ni yield similar to the yield obtained from the spherical explosion.
The $^{56}$Ni yield increases with the opening angle in 
$\theta_{op} \ga 45^\circ$ for 110 M$_\odot$ model and in 
$45^\circ \la \theta_{op} \la 68^\circ$ for 250 M$_\odot$ model.
The dependence on the opening angle is mainly due to the difference in the efficiency of
the fallback of the burned materials into the central remnant.
The amount of $^{56}$Ni produced during the explosion scarcely depends on the opening
angle.

The opening-angle dependence of the $^{28}$Si yield is similar to the $^{56}$Ni yield.
We see the increase with the opening angle in the $^{28}$Si yield 
for slightly smaller opening angles than in the $^{56}$Ni yield;
$45^\circ \la \theta_{op} \la 68^\circ$ for 110 M$_\odot$ model and 
$11^\circ \la \theta_{op} \la 68^\circ$ for 250 M$_\odot$ model.
This is mainly because Si is produced outside the $^{56}$Ni
forming region.
In larger opening angles, the fallback is suppressed in the Si forming region.
We see small dependence of the O yield on the opening angle.
This is because most of the O is located in outer regions.
A small amount of O in the equatorial region falls back and that in the polar region
burns into Si and other intermediate nuclei in small opening angle cases.

The total ejecta mass increases with an opening angle.
The lower limits of the ejecta mass are 28.4 M$_\odot$ and 47.5 M$_\odot$.
When the opening angle is small, the innermost materials fall back to the central remnant
through the equatorial region.
On the other hand, the explosion with a larger opening angle suppresses
 the fallback.
The aspherical explosion induces upstream along the polar axis as well as fallback.

We note that we simplified the equation of state and did not consider
the generation and absorption of nuclear energy with $\alpha$-network
in 2D explosion model.
So, we briefly discuss differences between the abundance distributions calculated
using 1D code and those using 2D code for spherical explosion models.
The difference of the $^{56}$Ni yield between 1D and 2D codes is less than 9\%.
For the yields of C, O, Na, Mg, Si, and Ca, which are listed in Table 3, 
the difference is less than 13 \% for Na, 12 \% for Ca, and 7\% for the other elements. 
Thus, although the $^{56}$Ni yield is slightly smaller in the result with the 2D code, 
we consider that difference in the adopting codes for solving the explosive 
nucleosynthesis does not give a large difference in the yields listed above.


\begin{figure*}
\includegraphics[width=6cm,angle=270]{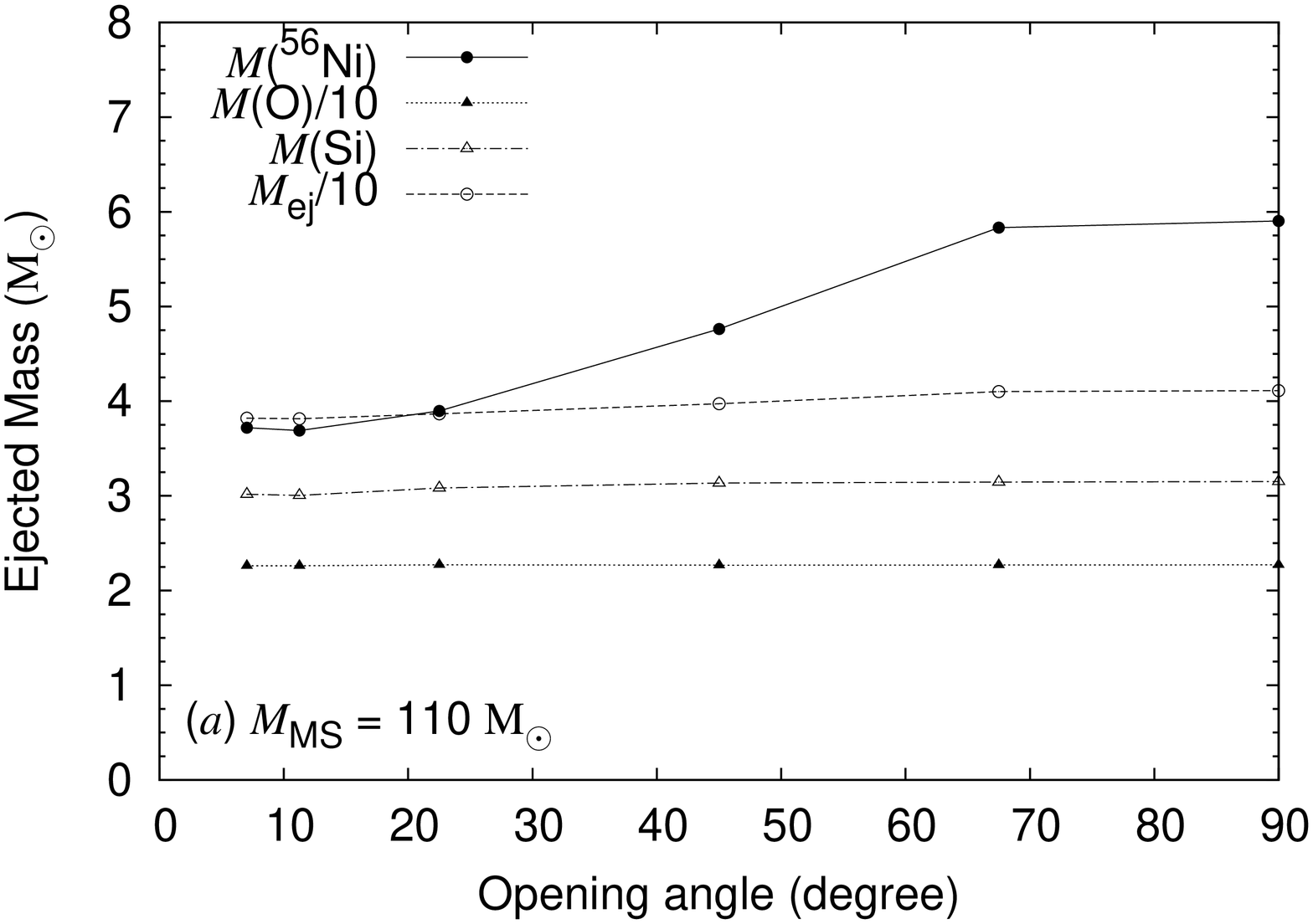}
\includegraphics[width=6cm,angle=270]{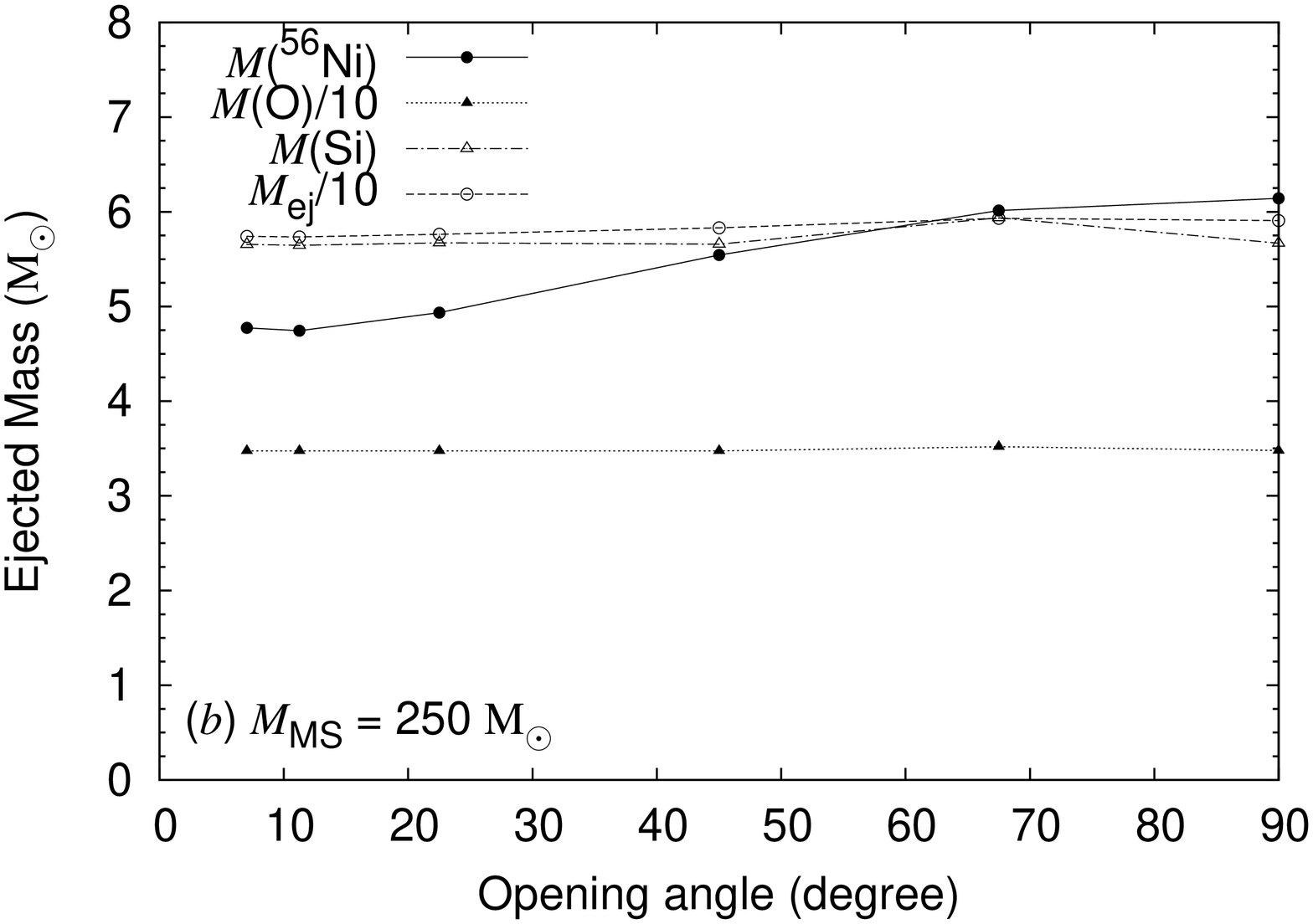}
\caption{Same as Fig. 4, but for the aspherical explosion models of
the collapsed progenitors.
}
\end{figure*}

\subsection{Aspherical explosions from collapsed progenitors}

Next, we evaluate the yields of the aspherical SN explosions from the collapsed progenitors.
Figure 5 shows the yields of $^{56}$Ni, O, Si and the total ejecta amount relating to the
opening angle of aspherical explosions with the collapsed 110 and 250 M$_\odot$ models.
The $^{56}$Ni yield is in the range of $3.7 - 5.9$ M$_\odot$ and 
$4.8-6.1$ M$_\odot$ for the 110 and 250 M$_\odot$ models, respectively.
Although the yields and the total ejecta mass increase with the opening angle, 
the dependence is much smaller than that of the corresponding uncollapsed model.
The yields of Si and O scarcely depend on the opening angle.
These yields are also very close to the corresponding spherical explosion models from 
the uncollapsed progenitors.

The opening-angle dependence of the yields in the collapsed models is very small.
The explosion of a small opening angle of a collapsed model indicated an explosion feature
close to spherical symmetry compared with the corresponding explosion of the uncollapsed
model.
The final spacial distribution of the yields of the collapsed model of 
$\theta_{op} = 11.25^\circ$
is close to the uncollapsed model of $\theta_{op} = 67.5^\circ$.
The ejecta mass also scarcely depends on the opening angle and is larger than that of 
the corresponding uncollapsed model.
In the collapsed model, 
the energy-injection radius is smaller by about one order of magnitude than that of the 
uncollapsed model.
The explosion becomes close to ^^ ^^ point-like" explosion even in $^{56}$Ni forming region.
Thus, the dependence on the opening angle is smaller in the collapsed models.

We note that we set the energy deposition time to be $10^{-3}$ s in this study.
However, energy deposition time of aspherical SNe is quite uncertain will strongly affect
the explosion feature and the yield distribution.
Larger energy deposition time is expected to reduce the production and ejection
of $^{56}$Ni \citep[e.g.][]{Tominaga07}.
The $^{56}$Ni yield evaluated using the collapsed model would be the maximum value of
the $^{56}$Ni yield with a given explosion energy and an opening angle.

\section{Discussion}

\subsection{SL SN from very massive stars}

Radioactive decay of $^{56}$Ni and $^{56}$Co is one of main light sources of SLSNe.
\citet{Umeda08} showed that more than 1 M$_\odot$ $^{56}$Ni is ejected by CC explosion
when the CO core is roughly larger than 10 M$_\odot$ and
the explosion energy is larger than $E_{{\rm ex,51}} = 10$.
A CO core larger than $\sim 35$ M$_\odot$ with 
$E_{{\rm ex,51}} \ga 20$ will eject more than 3 M$_\odot$ $^{56}$Ni (see also YU11).
On the other hand, a progenitor with a CO core larger than $\sim 60$ M$_\odot$ will explode as 
a PI SN.
We draw the shaded regions satisfying the above conditions on the CO core masses in Fig. 1($b$).
Therefore, we expect that CC SNe Ic which eject more than 1 M$_\odot$ $^{56}$Ni 
is possible for stars with $M_{{\rm MS}} \ga 30$ M$_\odot$ and $Z > 0.004$.
Core-collapse explosions evolved from very massive stars with 
$100 \la M_{{\rm MS}} \la 250$ M$_\odot$ and $0.001 < Z \la 0.004$ would become SLSNe Ic 
similar to SN 2007bi.
Stars with $M_{{\rm MS}} \sim 110 - 150$ M$_\odot$ and $Z \la 0.001$ are expected to experience
PPI and the whole H and He layer may be lost.
In this case, these stars will also explode as SLSNe Ic.

Recently, \citet{Yusof13} studied the evolution and fate of very massive stars.
They evaluated the initial mass range of a PI SN progenitor for SN 2007bi as
$M_{{\rm MS}} = 160 - 170$ M$_\odot$ based on their evolution calculations of 
$150 - 300$ M$_\odot$ rotating stars with $Z=0.002$.
This mass is smaller than the one evaluated in YU11.
One reason is the use of different mass loss rate.
The mass loss rate in WR phase in our study seems to be larger than their rate.
The metallicity dependence of our rate in MS stars ($\propto$ ($Z$/Z$_\odot$)$^{0.64-0.69}$) is
weaker than their rate ($\propto$ ($Z$/Z$_\odot$)$^{0.85}$) and thus, our rate is practically larger.
Another reason is larger CO-core mass in their models.
The CO-core masses of their $M_{{\rm MS}} = 150 - 200$ M$_\odot$ models are larger than
those of our models with $Z=10^{-4}$, in which mass loss does not affect the CO-core mass.
This is probably due to rotation-induced mixing.
Decrease in the lowest mass for PI SN by stellar rotation was shown in \citet{Chatzopoulos12a}.

\begin{table*}
\caption{
The yields of spherical and aspherical CC SN models and PI SN models.
In spherical CC SN models, M110E50 and M250E70 indicate the SN explosions of
($M_{{\rm MS}}$, $E_{{\rm ex,51}}$) = (110 M$_{\odot}$, 50) and
(250 M$_{\odot}$, 70), respectively.
In aspherical CC SN models, M110$\theta_{op}$11.25 and 
M250$\theta_{op}$11.25 indicate
the SN explosions of ($M_{{\rm MS}}$, $E_{{\rm ex,51}}$, $\theta_{op}$) = 
(110M$_\odot$, 50, 11.25$^\circ$) and 
(250M$_\odot$, 70, 11.25$^\circ$), respectively.
We adopted the results of the collapsed aspherical models.
For PI SN models, MHe100 and M200 correspond to the PI SN models of 
$M_{{\rm He}} = 100$ M$_\odot$ He star in \citet{Heger02}
and $M = 200$ M$_\odot$ metal-free star in \citet{Umeda02}.
SN 2007bi indicates the yield ranges of SN 2007bi obtained by six measurements 
listed in Table 1 of  \citet{Gal-Yam09}.
For $^{56}$Ni yield, the estimate from the observed peak luminosity is also taken into account.
}
\begin{tabular}{lccccccc}
\hline
 & \multicolumn{2}{c}{Spherical CC SN models} & 
\multicolumn{2}{c}{Aspherical CC SN models} &
\multicolumn{2}{c}{PI SN models} & SN 2007bi \\
Element & M110E50 & M250E70 &
M110$\theta_{op}$11.25 & M250$\theta_{op}$11.25 & 
MHe100 (HW02) & M200 (UN02) & \\
\hline
C         & 2.51  & 2.71 & 2.52 & 2.74 & 4.01  & 4.24 & $1.0-1.2$  \\
O         & 23.6 & 34.9 & 22.4 & 34.8  & 43.9  & 56.0 & $7.5-14.6$  \\
Na       & 000800 & 0.00831 & 0.00768 & 0.00852 & 0.00276 & 0.00700 & $0.0012-0.0023$ \\
Mg      & 1.10  & 1.53 & 1.01 & 1.56 & 4.41  & 3.08 & $0.065-0.13$  \\
Si        & 3.01 & 5.70 & 3.00 & 5.65 & 23.1  & 21.2  & --- \\
Ca      & 0.29 & 0.43 & 0.24 & 0.39  & 1.22  & 2.32  & $0.75-1.10$ \\
$^{56}$Ni & 5.13 & 6.11 & 3.69  & 4.74  & 5.82  & 7.2 & $3.5-7.4$ \\
\hline
\end{tabular}
\end{table*}

\subsection{Yields of SN 2007bi}

SN 2007bi is considered to eject $^{56}$Ni of $3.5 - 7.4$ M$_\odot$.
In order to explain the $^{56}$Ni amount by spherical SN explosion of a $40 - 60$ M$_\odot$
CO star, the explosion energy should be $E_{{\rm ex,51}} \ga 20$.
The explosion should be very energetic.
This result is consistent with the estimate of the ejected $^{56}$Ni amount in \citet{Umeda08}.
On the other hand, if the explosion energy is $E_{{\rm ex,51}} \la 5$, almost all $^{56}$Ni
falls-back into the central remnant even if $^{56}$Ni is produced explosively.
In such a case, the exploded materials cannot be brighten by the radioactive decays of $^{56}$Ni
and $^{56}$Co.
Normal explosion of a very massive star will become a faint SN.

Here we discuss a possibility of aspherical CC explosion of SN 2007bi.
In the case of the collapsed progenitor models, which is expected to be more realistic,
all aspherical explosions of the 110 and 250 M$_\odot$ models 
reproduce the $^{56}$Ni yield of SN 2007bi.
In the uncollapsed models, the range of the opening angle obtaining the $^{56}$Ni mass
more than 3.5 M$_\odot$ is $\theta_{op} \ga 68^\circ$ and 35$^\circ$ for the 110 and
250 M$_\odot$ models, respectively.
Thus, aspherical explosions with a small opening angle would eject the $^{56}$Ni amount enough to reproduce the $^{56}$Ni yield of SN 2007bi,
if the central region of the progenitor has collapsed before the explosion.
We expect that aspherical CC SN explosion with a $\sim 40 - 60$ M$_\odot$ CO-core progenitor
ejects $^{56}$Ni enough to reproduce the $^{56}$Ni yield observed in SN 2007bi.
The degree of asphericity of the explosion reproducing observational features of SN 2007bi
depends on progenitor mass and explosion energy.


The elemental yields of C, N, O, Na, Mg, and Ca were measured using late-time spectra in SN 2007bi
\citep{Gal-Yam09}.
We show the yields of these elements for four cases of the CC SN models in this study and PI SN
models in \citet{Heger02} and \citet{Umeda02} in Table 3.
The ranges of the measured yields in SN 2007bi are also listed.

For most elements, the yields in both the CC SN models and PI SN models are larger than
the corresponding observed yields.
On the other hand, the ejecta mass of SN 2007bi was estimated to be  
$36 < M_{{\rm ej}} < 173$ M$_\odot$ from the scaling relations of the rise time and 
the photospheric velocity \citep[Supplementary information of][]{Gal-Yam09}.
They suggested that a part of light elements are contained in unilluminated materials.
The hidden amount of the ejecta is still quite uncertain.
Therefore, we consider that it is still difficult to discuss the differences between the yields
evaluated in CC SN and PI SN models and the observed yields.
It is important to find the relation between the ejecta containing such a huge mass and the
emitted spectra.


\begin{figure*}
\includegraphics[width=6cm,angle=270]{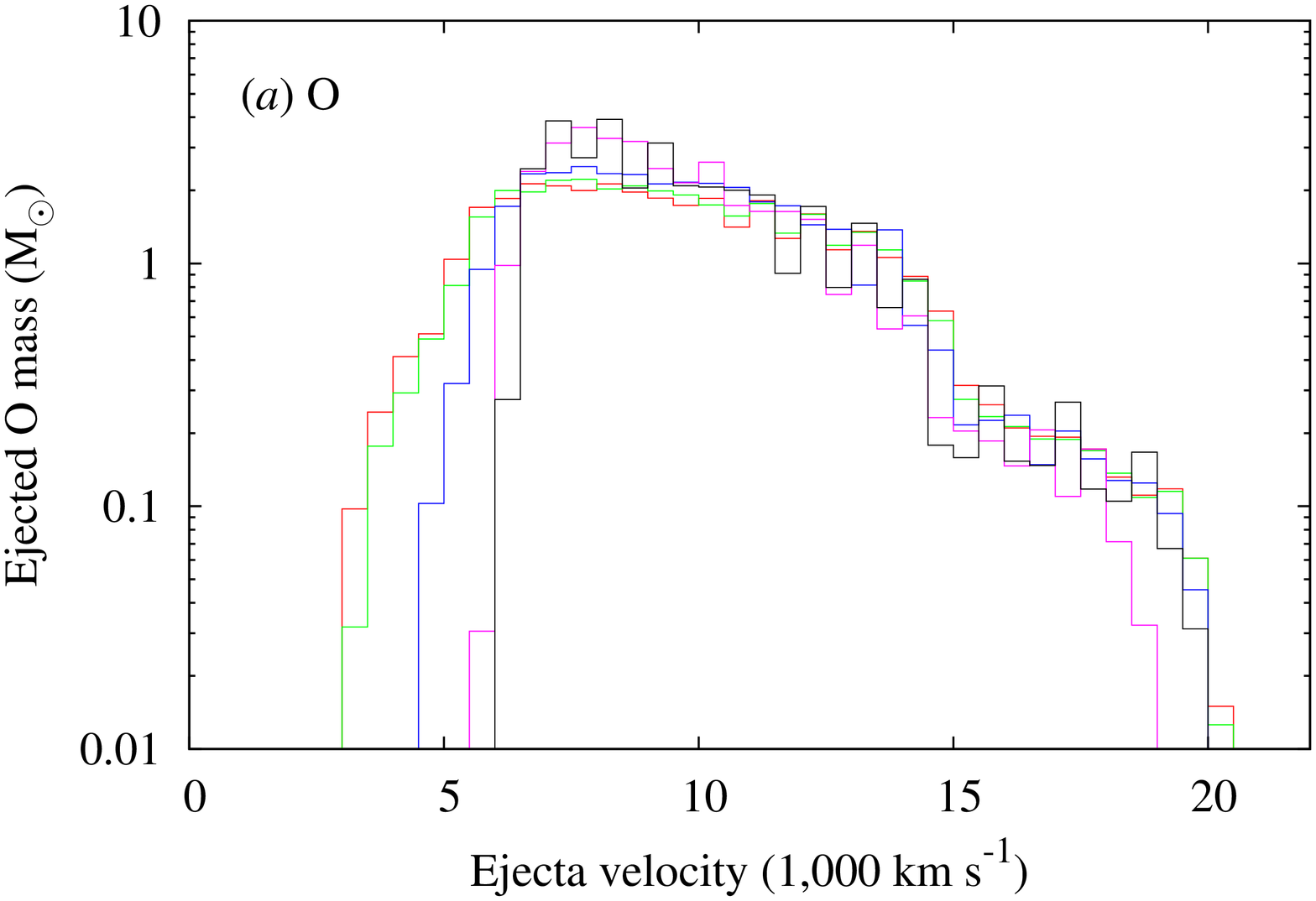}
\includegraphics[width=6cm,angle=270]{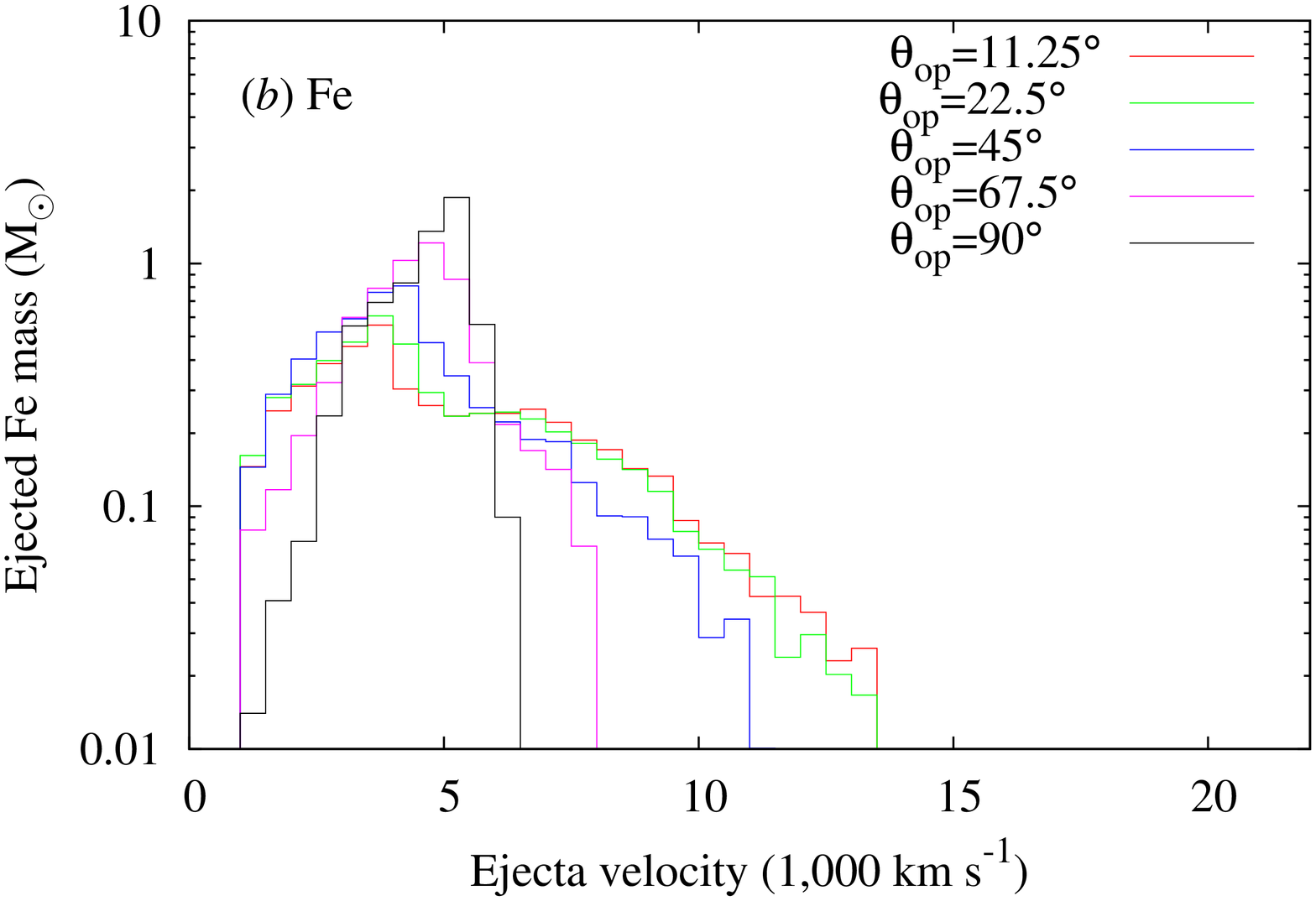}
\includegraphics[width=6cm,angle=270]{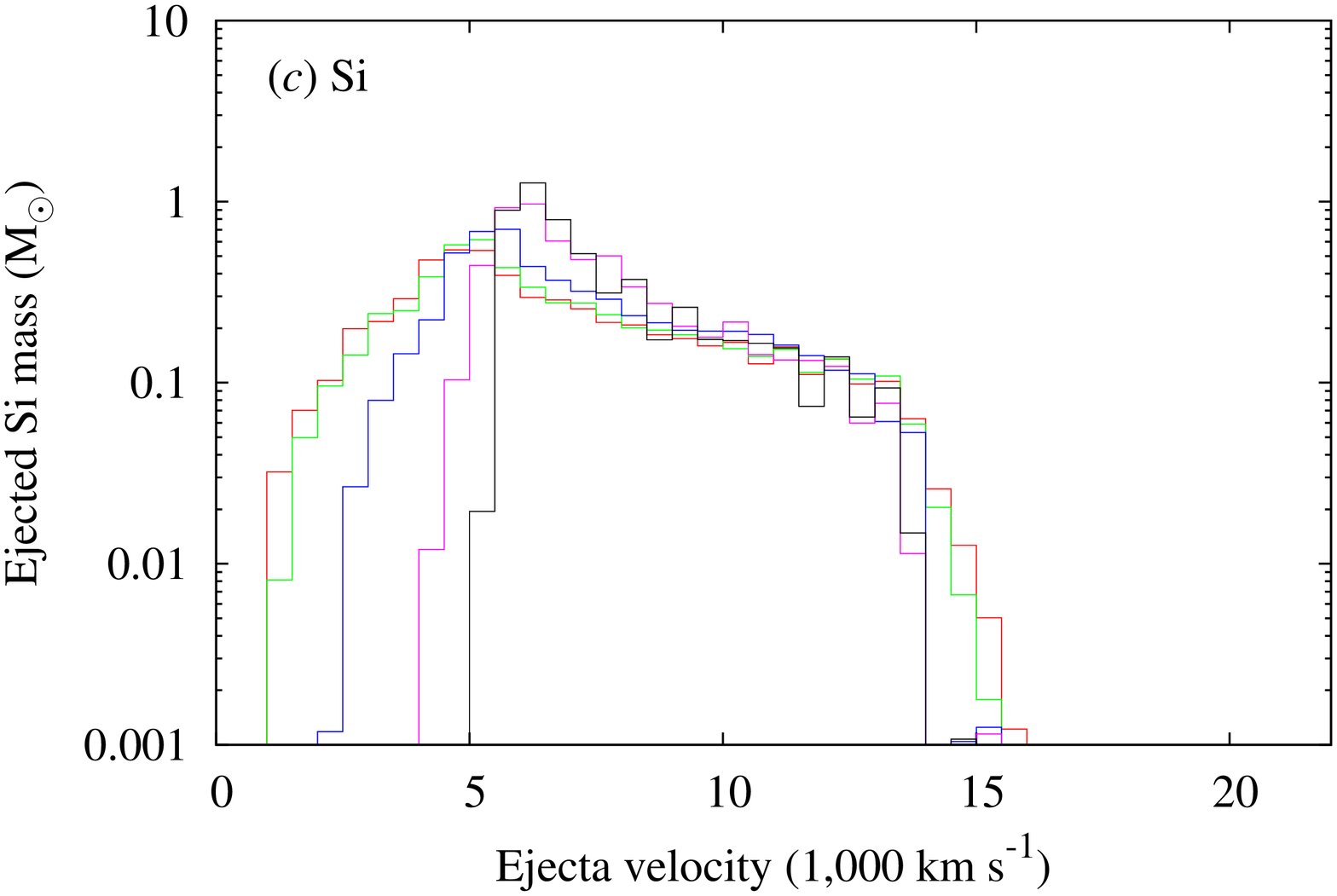}
\includegraphics[width=6cm,angle=270]{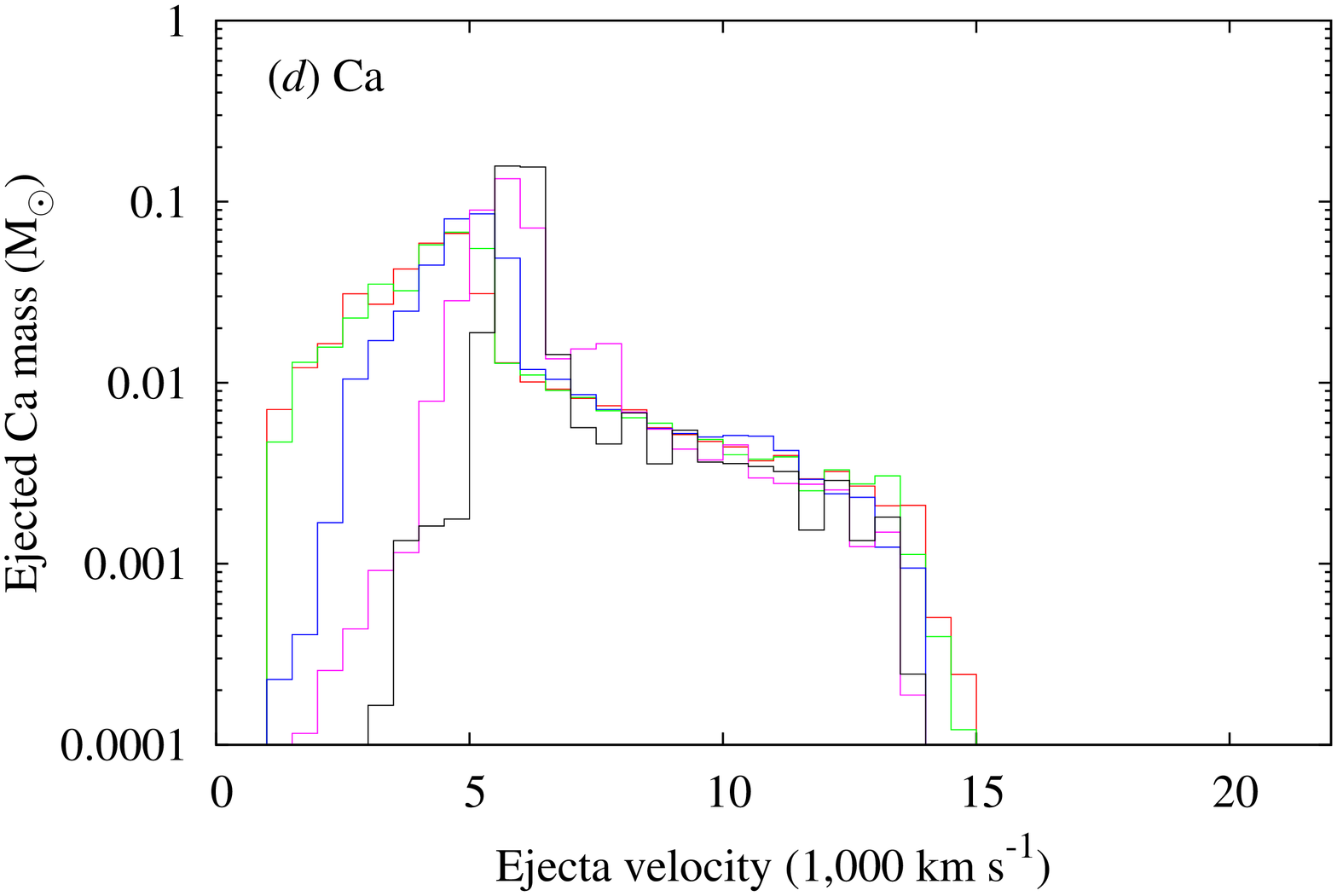}
\includegraphics[width=6cm,angle=270]{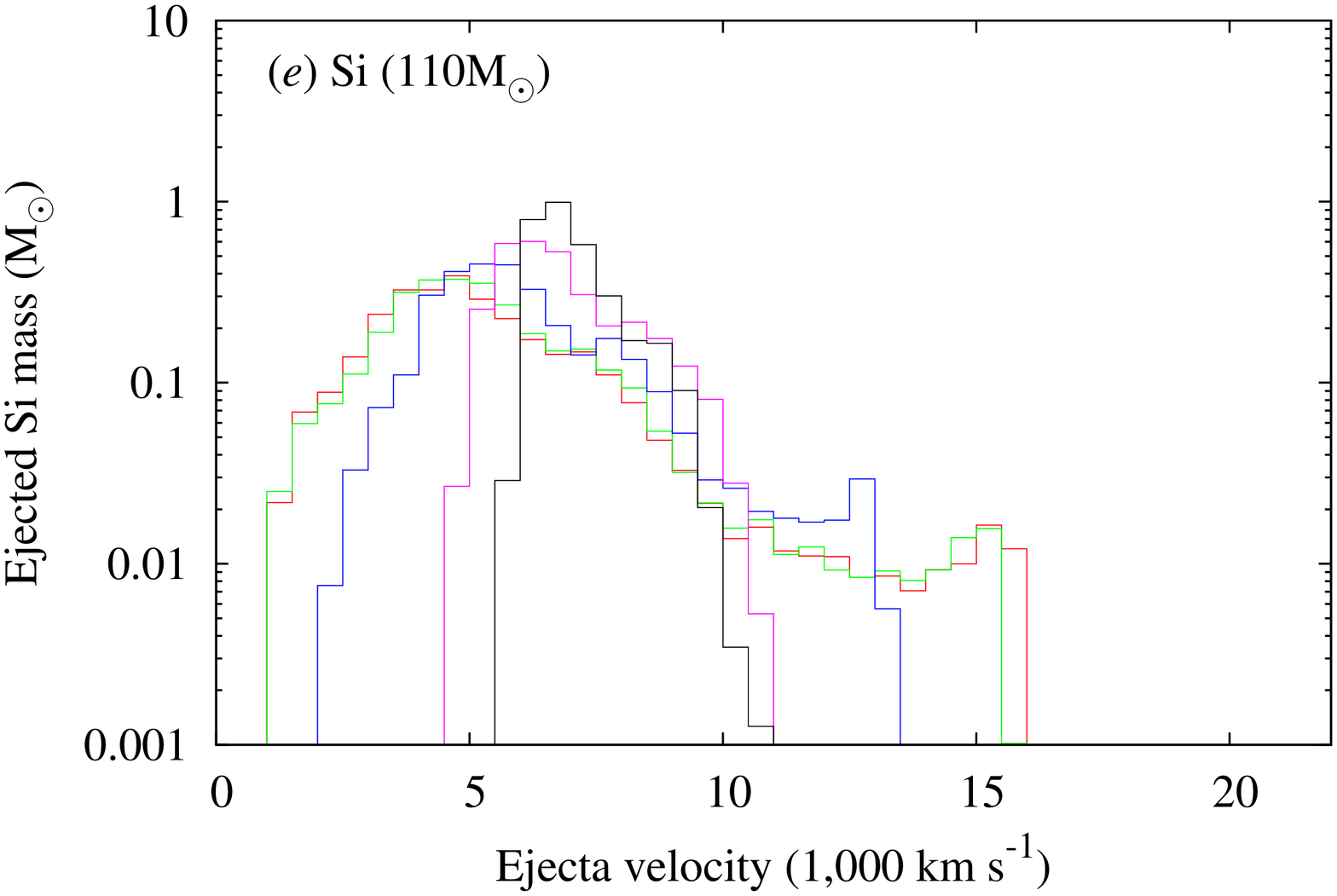}
\includegraphics[width=6cm,angle=270]{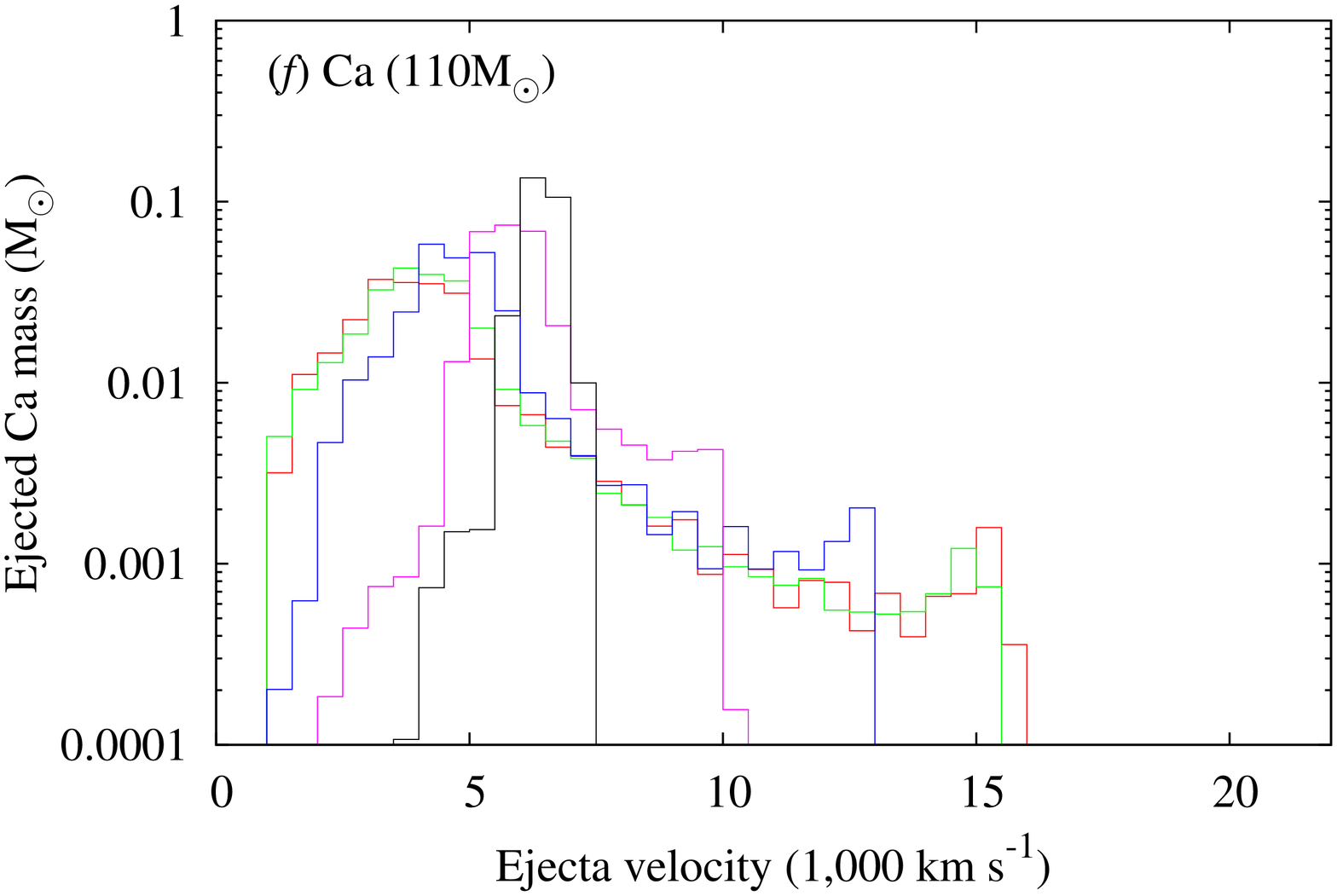}
\caption{Velocity distributions of ejected masses of O ($a$), Fe ($b$), 
Si ($c$), and Ca ($d$) in aspherical explosions of the collapsed 250 M$_\odot$ model, and 
Si ($e$) and Ca ($f$) in the collapsed 110 M$\odot$ model.
Red, green blue, purple, and black lines correspond to the cases of the opening angles of
$\theta_{op} = 11.25^\circ$, 22.5$^\circ$, 45$^\circ$, 67.25$^\circ$, and 90$^\circ$, respectively.
}
\end{figure*}

\subsection{Ejecta velocities and rise time of light curve}

The ejecta velocities of SN 2007bi were estimated by using the spectra of 
O, Fe, Si, and Ca
in $54-134$ days \citep{Young10}.
The obtained ejecta velocities are $12,000 - 16,000$ km s$^{-1}$ for Ca II
H \& K lines and $10,000 - 12,000$ km s$^{-1}$ for Fe II. 
$7,600-9,800$ km s$^{-1}$ for Ca II near-IR $5,500 - 8,000$ km s$^{-1}$ for Si II, and 4,500 km s 
$^{-1}$ for the lower limit of O I. 

Here, we evaluate the velocity distributions of O, Fe, Si, and Ca 
in aspherical explosions of 
the collapsed 110 and 250 M$_\odot$ models.
We note that the observed ejecta velocities do not indicate the bulk velocities of 
the ejecta. This is
because the observed period is before the nebular phase and the velocities depend on the
strength of absorption.
Thus, it is difficult to discuss explosion features from the comparison between 
the observed velocities and the evaluated ones.
Although we do not discuss the comparison between the observed and evaluated
velocities, this evaluation will help understanding explosion features of SLSNe that will be observed in the future through the observations of the ejecta velocities in the nebular
phase.

Figure 6 shows the ejected masses of O, Fe, Si, and Ca as a function of the ejecta velocity
and the opening angle.
Panels ($a$)-($d$) indicate the velocity distributions of the collapsed 250 M$_\odot$ model.
Most of O ejecta have a velocity of $\sim 6,000 - 20,000$ km s$^{-1}$ with small
opening-angle dependence.
The minimum velocity and the amount of slower ejecta indicate more significant dependence on the opening angle.
The minimum velocity is about 6,000 km s$^{-1}$ for spherical explosion and 
$\sim 3,000$ km s$^{-1}$ for the explosion with $\theta_{op} = 11.25^\circ$.
The amount of slower O ejecta increases with decreasing opening angle of
the explosion.

The velocity distribution of Fe ejecta also depends on the opening angle.
The velocity range of the Fe ejecta in the spherical explosion is $1,000 - 6,500$ km s$^{-1}$, 
which is narrower than O for the spherical explosion.
The velocity dispersion is caused by the large $^{56}$Ni production in wide spatial
range of the innermost region.
As the opening angle decreases, the maximum velocity becomes faster
and slow ejecta increases.
At the same time, the velocity dependence of the ejected amount becomes smaller.
In the case of $\theta_{op} = 11.25^\circ$, the velocity range extends to 
$1,000 - 13,500$ km s$^{-1}$ .
Both the ejecta masses of faster and slower components increase.

The velocity dependence of Si and Ca masses moderately depends on the opening angle.
The velocity range in the spherical explosion is $\sim 5,000 - 14,000$ km s$^{-1}$.
The minimum velocity decreases and slower ejecta increases with
decreasing the opening angle.

We also show the velocity distributions of Si and Ca masses of the 110 M$_\odot$ model
in Figs. 6 ($e$) and ($f$), respectively.
Although the velocity range is narrow in spherical explosion, the
dependence of the velocity on the opening angle is similar to the
250 M$_\odot$ model.
Thus, aspherical explosions of the collapsed models produce wide velocity
dispersion of SN ejecta.

We briefly discuss the velocity distributions in the uncollapsed models.
When the opening angle of the explosion is small, the velocity distribution is larger than the corresponding collapsed model.
In the case of the 250 M$_\odot$ model with $\theta_{op} = 11.25^\circ$, the O mass decreases
to 1,500 km s$^{-1}$ and the Fe, Si, and Ca masses increase up to 18,000 km s$^{-1}$.
In the case of the 110 M$_\odot$ model with $\theta_{op} = 11.25^\circ$, the fastest component
of Si and Ca is 20,000 km s$^{-1}$.
On the other hand, the velocity distributions of the explosion models reproducing the $^{56}$Ni mass of SN 2007bi are also similar to the spherical explosions
of the collapsed models.
These explosion models have moderate opening angles.

At last, spherical explosions indicated narrow ranges of the ejecta velocities for Fe
in the 110 and 250 M$_\odot$ models and Si and Ca in the 110 M$_\odot$ model.
On the other hand, aspherical explosions indicated wider ranges of the ejecta 
velocities.
An aspherical explosion causes wide velocity distribution of various elements even if
large-scale mixing does not occur in late time.
These characteristics of the velocity distribution would constrain sphericity of SLSNe.


The rise time is a parameter characterizing the light curve of a SN.
Spherical CC SN model in \citet{Moriya10} reproduced the light curve of SN 2007bi.
They evaluated the light curve assuming three cases of ejecta mixing and obtained the
rise time of 52 days (full mixing), 67 days (half mixing), and 85 days (without mixing).
They also reproduced the light curve by a PI SN model.
The obtained rise time is $\sim 150$ days.
The rise time can be approximately estimated as the relation 
$\tau_{{\rm rise}} \propto (M_{{\rm ej}}^3/E_{{\rm ex}})^{1/4}$ \citep[e.g.][]{Nakamura01}.
The rise time is roughly evaluated as 
$\tau_{{\rm rise}} \sim (7.6 - 12.4) (M_{{\rm ej}}^3/E_{{\rm ex,51}})^{1/4}$ days using
$(M_{{\rm ej}}^3/E_{{\rm ex,51}})^{1/4} = 6.85$ for the CC explosion model in \citet{Moriya10}.
Taking account of the uncertainties in the rise time and the mixing, we consider that
the range of $(M_{{\rm ej}}^3/E_{{\rm ex,51}})^{1/4}$ available for the rise time of SN 2007bi
is between 4.2 (short rise time and without mixing) and 11.2 (long rise time and full mixing).
Here we discuss the rise time of our SN models taking account this dependence.

In spherical explosions, we investigated the dependence on the explosion energy.
The rise time becomes shorter with increasing in the explosion energy.
When the explosion energy is between $E_{{\rm ex,51}} = 20$ and 100, where the $^{56}$Ni
yield of SN 2007bi is reproduced, the range of $(M_{{\rm ej}}^3/E_{{\rm ex,51}})^{1/4}$ is
$5.12 - 7.66$ and $6.74 - 10.1$ for the 110 and 250 M$_\odot$ models, respectively.
Thus, these explosion models would reproduce the light curve of SN 2007bi if ejecta mixing
is appropriate.

In aspherical explosions, we investigated the dependence on the opening angle with a given
explosion energy.
Since the ejecta mass increases with the opening angle, the rise time also increases.
The range of $(M_{{\rm ej}}^3/E_{{\rm ex,51}})^{1/4}$ is 
$5.77 (4.63) - 6.11 (6.08)$ and 
$7.21 (6.26) - 7.37 (7.36)$ for the collapsed (uncollapsed) 110 and 250 M$_\odot$ models.
The rise time of the 250 M$_\odot$ models seems to be close to that in the CC SN model
in \citet{Moriya10}.
On the other hand, the rise time of jet-like explosion of the 110 M$_\odot$ models would
be small.
Among the explosion models reproducing the $^{56}$Ni yield of SN 2007bi, the minimum value of
$(M_{{\rm ej}}^3/E_{{\rm ex,51}})^{1/4}$ is 5.77 (6.08) and 7.20 (6.92) 
for the collapsed (uncollapsed) 110 and 250 M$_\odot$ models.
Thus, we expect that spherical and aspherical CC explosion models reproducing 
the $^{56}$Ni yield of SN 2007bi would also reproduce the light curve of SN 2007bi.

\subsection{Surface He amount}

In our progenitor models, a small amount, less than 1 M$_\odot$, of He remains 
in the outer layer.
PI SN models for SN 2007bi suggested in \citet{Yusof13} would also remain He in the envelope.
The He amount hidden in an SN Ic is an unclarified problem.
He lines appear through the excitation of He by non-thermal electrons
induced by $\gamma$-rays from $^{56}$Ni and $^{56}$Co.
The strength of the lines complexly depends on the amounts of He and $^{56}$Ni
and other elements to thermalize electrons.
The He amount hidden in $1-3$ M$_\odot$ SN Ic ejecta was evaluated in
\citet{Hachinger12}.
The He amount hidden in more massive SNe Ic should be investigated
for discussing the possibilities of SNe Ic from metal-poor very massive 
stars and PI SNe Ic.



\subsection{Mass loss during pulsational pair-instability}

A star with $M_{{\rm MS}} = 250$ M$_\odot$ experienced PPI during the Si burning.
Although eruptive mass loss is induced by PPI, we did not consider this effect.
The mass lost during PPI for the He star models with $48 - 60$ M$_\odot$ was investigated in
\cite{Woosley07} (see their supplementary information).
They obtained that the mass of $7 - 18$ M$_\odot$ was lost during PPI and the final mass ranges
in $40 - 49$ M$_\odot$.
\cite{Chatzopoulos12} evaluated the mass lost in one pulsation of PPI for metal-poor rotating
stars.
The pre-PPI SN stars with $41 - 58$ M$_\odot$ lost the mass of $1.9 - 7.3$ M$_\odot$.
Thus, we expect that the 250 M$_\odot$ star in this study will become $\sim 40 - 50$ M$_\odot$ 
after PPI.
In this case, explosion feature of the SN will be similar to the SN explosion of the 110 M$_\odot$
model (43 M$_\odot$ progenitor) and aspherical explosion reproducing the $^{56}$Ni yield
of SN 2007bi would be possible.
The above two studies also discussed the possibility that the collisions of the ejected material with 
the earlier ejecta during PPI induce very bright events like SLSNe.
Recently, analytical light curve models indicated that the light curve of SN 2007bi can be 
explained by the collision of SN ejecta with hydrogen-deficient circumstellar matter
in addition to the radioactive decays of $^{56}$Ni and $^{56}$Co of $\sim 0.5$ M$_\odot$
\citep{Chatzopoulos13}.
In this case, less energetic explosion is possible.

\subsection{Prospects}

We set the explosion energy of $E_{{\rm ex,51}} = 50$ and 70 for the 110 M$_\odot$ and 
250 M$_\odot$ SN models, respectively.
Such energetic explosions have been estimated through observations in Type Ic SN 1999as
and GRB 031203/SN 2003lw \citep[e.g.,][]{Nomoto06}.
There are some possibilities of the explosion mechanism of aspherical CC SNe.
One is a launch of the relativistic jet driven by magnetar 
formation\citep[e.g.,][]{Takiwaki09}.
Another is a black-hole and accretion disk formation and the consequent jet production
, i.e., collapsar scenario\citep{MacFadyen99}.
These mechanisms would have induced an aspherical energetic explosion and the explosion
would have produced the $^{56}$Ni yield observed in SN 2007bi.

SNe Ic associated with gamma-ray bursts (GRB/SNe Ic) have been found 
in host galaxies having similar metallicities to the host galaxy of SN 2007bi \citep{Young10}.
GRB is a highly jet-like event and GRB/SNe Ic were observed as jet-like explosions \citep{Maeda08}.
Broad-lined Type Ic SNe also have been found in slightly metal-richer environments than the
host galaxies of GRB/SNe Ic.
Discussion of the relation between these SNe Ic and SLSNe Ic like SN 2007bi would be important
for clarifying the SN events during galactic chemical evolution.

We should note that such an extreme energetic CC explosion has not been 
observationally ruled out, although the explosion mechanism has not been theoretically 
established.
We showed in this study the possibility of such an extreme energetic CC explosion 
enough to explain the $^{56}$Ni yield of SN 2007bi within uncertainties in the 
mass loss rate in very massive stars.
As a significance of this study, we consider that this study will extend to future studies 
to clarify if such an explosion is possible or not.

There are several SLSNe expected to be the explosions from very massive stars.
SN 1999as showed the light curve similar to SN 2007bi in photospheric phase \citep{Gal-Yam12}
and more than 4 M$_\odot$ of $^{56}$Ni was estimated to be ejected \citep{Deng01}.
Recently, PTF10nmn observed by the Palomar Transient Factory and PS1-11ap found by 
the Panoramic Survey Telescope and Rapid Response System 1 were suggested to be
candidates of PI SNe \citep[][and references therein]{Gal-Yam12}.
Additional two SLSNe were also observed at redshifts $z=2.05$ and 3.90.
They are also candidates of a PI SN and a PPI SN from photometric and far-ultraviolet data
\citep{Cooke12}.
Very recently, the observations of PTF12dam and PS1-11ap were reported and 
magnetar-energized ejecta were proposed \citep{Nicholl13}.
Although the event rate of such SLSNe is very small, the measurements of the SLSNe, which
help constraining explosion mechanism and progenitor mass of each SN, will increase in
future.

\section{Conclusions}

We have investigated the dependence of the final mass and the fate
on the MS mass and the metallicity of massive stars.
We expect that very massive stars with $M_{{\rm MS}} \ga 100$ M$_\odot$ and
$0.001 < Z \la 0.004$ become CC SLSNe Ic like SN 2007bi.
Metal-poorer stars would explode as SNe II or SNe Ib even if they explode
as PI SNe.
Stars with $M_{{\rm MS}} \sim 110 - 150$ M$_\odot$ and $Z \la 0.001$ would become
PPI SNe Ic if the whole H and He layers is lost during PPI.
The evaluation of the He amount in SN ejecta of very massive stars
to make He lines is important for determining SN types.

We also investigated the dependence of the total ejecta mass
and the yields of $^{56}$Ni, O, and Si on the explosion 
energy as well as asphericity of the CC SN models
with $M_{\rm MS} = 110$ M$_\odot$ and 250 M$_\odot$ and $Z=0.004$.
In the case of spherical explosions, the yield of $^{56}$Ni produced in 
the SN increases with the explosion energy.
The SN explosion with $E_{{\rm ex},51} \ga 20$ produces the $^{56}$Ni amount
enough to reproduce the amount observed in SN 2007bi.
In the case of aspherical explosions, the total ejecta mass and the yields of $^{56}$Ni, O, and Si 
increase with the opening angle.
The aspherical CC SNe of the collapsed 110 M$_\odot$ and 250 M$_\odot$ models reproduce the $^{56}$Ni yield observed in SN 2007bi.
In the uncollapsed progenitor models, moderately aspherical CC SNe also reproduce.
These SNe indicate the velocity distribution up to $\sim 13,000 - 15,000$ km s$^{-1}$ in
O, Si, Ca, and Fe.
Therefore, an aspherical CC SN explosion evolved from a very massive star is a possibility for the 
explosion of SN 2007bi.
The relation between SLSNe like SN 2007bi and GRB/SNe Ic and broad-lines SNe Ic would 
bring about new knowledge of SN events during galactic chemical evolution.

\section*{Acknowledgments}

We thank anonymous referee for giving us valuable comments.
We thank Hideyuki Saio for providing the stellar evolution code
and useful comments.
We are grateful to Masaomi Tanaka, Nobuyuki Iwamoto, Ken'ichi Nomoto, Takashi Moriya, 
and Koh Takahashi for valuable discussions.
We are indebted to Hamid Hamidani and Aaron C. Bell for reading our manuscript and giving variable comments.
This work was supported by the Grants-in-Aid for Scientific Research
(20041005, 20105004, 23540287, 24244028).

\bsp

\label{lastpage}

\end{document}